\documentclass[aps,prl,reprint,superscriptaddress]{revtex4-1}
\usepackage{graphicx}
\usepackage{color}

\usepackage{hyperref}
\usepackage{rotating}
\usepackage{booktabs}
\usepackage{multirow}

\begin{document}



\title{Accelerating  material melting temperature predictions \\by implementing 
 machine learning potentials in the \texttt {SLUSCHI} package}


\author{Audrey CampBell}
\affiliation{School for Engineering of Matter, Transport, and Energy, Arizona State University, Tempe, Arizona 85285, USA}

\author{Ligen Wang}
\affiliation{School for Engineering of Matter, Transport, and Energy, Arizona State University, Tempe, Arizona 85285, USA}



\author{Qi-Jun Hong}
\email[]{qijun.hong@asu.edu}
\affiliation{School for Engineering of Matter, Transport, and Energy, Arizona State University, Tempe, Arizona 85285, USA}


\date{\today}

\begin{abstract}
The \texttt {SLUSCHI} (Solid and Liquid in Ultra Small Coexistence with Hovering Interfaces) automated package, with interface to the first-principles code \texttt{VASP} (Vienna Ab initio Simulation Package), was developed by us \cite{hong_user_2016} for efficiently determining the melting temperatures of various materials. However, performing many \texttt{DFT} molecular dynamics simulations for small liquid-solid coexisting supercells to predict the melting temperature of a material is still computationally expensive, often requiring weeks and tens to hundreds of thousands of CPU hours to complete. In the present paper, we made an attempt to interface the SLUSCHI package with the highly efficient molecular dynamics \texttt{LAMMPS} code and  demonstrated that it achieves a much faster melting temperature determination, outperforming the original  \texttt{VASP}-based approach by at least one order of magnitude. In our melting temperature calculations, the \texttt{LAMMPS} simulations were performed based on the \texttt{LASP} (Large-scale Atomic Simulation) machine learning potentials which are pre-built using first-principles data.  Besides the dramatic CPU time reduction for melting temperature predictions the calculated melting temperatures for various materials (simple and transition metals, alloys, oxides and carbide) are reasonably accurate. Analysis of the calculated results shows that 60\% of the melting temperatures are within 200 K of experimental values with the RMSE value of 187 K which is slightly worse than the first-principles \texttt{DFT} RMSE value of 151 K.  Therefore, interfacing \texttt{SLUSCHI} with \texttt{LAMMPS} molecular dynamics simulations makes it possible to quickly screen out the best candidates from numerous materials in a much more efficient way, and facilitate the rational design of materials within the framework of the materials genome paradigm.

\pacs{2}


\end{abstract}



\maketitle




\section{ \textbf{Introduction} }
Finding ways to determine the melting temperature of materials is a field of research that continues to interest materials researchers. From an experimental stand point, new and improved testing methods allow for materials to be tested under unique environments. Therefore, there is a corresponding need to improve the computational ability to calculate the melting temperature. There exists a number of already proven methods for calculating the melting temperature, including: the free energy method \cite{Free_energy_3,Free_energy_4,Free_energy_5,Free_energy_method_1,Free_energy_method_2}, the traditional large-scale coexistence method \cite{large_scale_1,large_scale_2} , the fast heating method \cite{Fast_heating_1,Fast_heating_2,Fast_heating_3,Fast_heating_4}, and the solid and liquid in ultra small coexistence (\texttt{SLUSCHI}) method \cite{SLUSCHI_table_1, SLUSCHI_table_2,SLUSCHI_table_3,SLUSCHI_table_4,hong_user_2016} . 

In the free energy method \cite{Free_energy_3,Free_energy_4,Free_energy_5,Free_energy_method_1,Free_energy_method_2} the melting temperature is determined by analyzing the intersection  of free energy curves calculated over a range of temperatures. The melting temperature is defined by the intersection between the solid and liquid melting curves where the free energy of both the liquid and solid exist at the same energy and temperature. However, for some materials the intersection between the two curves is less clear, due to the low angle of intersection, caused by the two curves having similar energies at the melting point. 

The traditional large-scale coexistence method, relies on identifying stable coexistence through \texttt{NPT} (constant number of particles, pressure, and temperature) molecular dynamics simulations, by directly simulating the interface between the liquid and solid phases.  However, its efficacy is contingent upon employing a large system size, which subsequently increases the computational requirements.\cite{large_scale_1,large_scale_2} 

The fast-heating method involves incrementally raising the temperature until melting occurs, employing \texttt{NVE} (constant number of particles, volume, and energy) molecular dynamics simulations. Through this method it is common to exceed the actual melting temperature because the system requires more energy to overcome the phase change, which in turn results in a hysteresis reaction. Therefore, this method requires corrections to achieve a higher level of precision when applied to larger systems or longer simulations.\cite{Fast_heating_1,Fast_heating_2,Fast_heating_3,Fast_heating_4} 

The solid and liquid in ultra small coexistence with hovering interfaces, \texttt{SLUSCHI} \cite{SLUSCHI_table_1, SLUSCHI_table_2,SLUSCHI_table_3,SLUSCHI_table_4,hong_user_2016} method was developed to solve the problems mentioned in the previous three methods. \texttt{SLUSCHI}, is a computational package developed for calculating the solid-liquid phase boundaries via first principles methods. The code uses smaller sample sizes which in turn reduces the computational time needed to run the simulations. It address the instability traditionally associated with small simulation sizes due to the thermal fluctuations at the interface by deploying multiple calculations to determine the coexistence of the solid and liquid phases at the interface allowing for a statistical analysis of the fluctuations thereby limiting the errors associated with the instability of the interface. Furthermore, \texttt{SLUSCHI} can be fully automated, minimizing the need for extensive human-computer interaction during the simulation process. 

The simulations are more streamlined which expedites the computational workflow, enhancing efficiency and reducing user intervention. Furthermore, \texttt{SLUSCHI} traditionally leverages density functional theory (DFT) and can interface with the first principle \texttt{VASP} code to calculate the melting temperature. By utilizing DFT, \texttt{SLUSCHI} employs a robust theoretical framework that accounts for electronic structure effects, leading to a more accurate depiction of the solid-liquid phase boundary, and accurate melting temperature predictions.\cite{hong_user_2016}

Although \texttt{SLUSCHI} is a highly efficient package for determining the melting temperature of a material, performing many coexisting molecular dynamics \textit{DFT AIMD} simulations is still computationally expensive. As shown in a previous publication \cite{hong_user_2016} it may take weeks to months and tens to hundreds of thousands of CPU hours to obtain the melting temperature for a material. In order to reduce the computational cost for determining the melting temperature of a material the \texttt{SLUSCHI} package is modified to interface with the highly efficient molecular dynamics package, \texttt{LAMMPS} (Large-scale Atomic/Molecular Massively Parallel Simulator). \texttt{LASP}'s database of pre-built machine learning potentials \cite{lasp_1,lasp_2}, which are based on  the results of first-principles calculations, were employed to perform  \texttt{LAMMPS} simulations. This approach eliminated the need for DFT molecular dynamics simulations when determining the melting temperature, which in turn leads to a dramatic reduction in computational costs. The CPU time results indicate that it outperforms the original \texttt{VASP}-based approach by at least one order of magnitude and the calculated melting temperatures for various materials (simple and transition metals, alloys and oxides) are still reasonably accurate.

\section{methods}

In the melting temperature calculation by \texttt{SLUSCHI}, many isobaric–isothermal (NPT) MD simulations are run at various temperatures for the solid–liquid coexisting supercell, allowing the supercell to evolve into a single solid or liquid phase. These first-principles MD simulations are quite expensive and it can take  tens to hundreds of thousands of CPU hours to carry out the \texttt{SLUSCHI} calculation for the melting point of a material. 
In this study, in order to speed up melting temperature prediction the \texttt{SLUSCHI} package is made to interface with highly efficient and well developed molecular dynamics program \texttt{LAMMPS}. 
The machine learning \texttt{LASP} potentials, pre-built from \texttt{VASP} calculated data, are employed for performing the \texttt{LAMMPS} simulations.

The \texttt{LASP} software was developed in the Liu group \cite{lasp_1,lasp_2} by integrating the neural network potential technique with the
stochastic surface walking  method. The \texttt{LASP} potentials are constructed by fitting to the data (energies, forces and stresses) of first-principles calculations.  A versatile machine learning  \texttt{LASP} potential database caters to a broad spectrum of computational tasks, such as structure prediction and reaction mechanism exploration. It has been implemented machine learning (ML) interatomic potentials in the\texttt{LAMMPS} (Large-scale Atomic/Molecular Massively Parallel Simulator) molecular dynamics code, and thus is ready for running \texttt{LAMMPS} simulations for melting temperature predictions with the \texttt{SLUSCHI} package.

To get the \texttt{LAMMPS} software to run the \texttt{LASP} potential \texttt{LAMMPS} needs to be compiled with the \texttt{LASP} settings, and the potential type in the \texttt{LAMMPS} input file has to be set to  the \texttt{LASP} potential type.  All the other tags can be adjusted for the simulation as normal for a \texttt{LAMMPS} run. These changes enable a streamlined approach to MD simulations and melting temperature determinations that leverages the capabilities of \texttt{SLUSCHI} for initial setup and integration into the simulation workflow.

During the simulation the initial atoms positions for the supercell are constructed following the same procedure as used in the \texttt{SLUSCHI} calculations with \texttt{VASP}. The unit cell for a material is optimized first using \texttt{VASP} with PBE exchange–correlation functional, then the supercell is constructed using the optimized unit cell according to the specified radius. The solid-liquid coexistence supercell is generated by duplicating the supercell such that the resulting supercell is doubled. The interface is thus created between the two supercells. 

Following the creation of the doubled supercell half of the resulting supercell is heated to melting, and  \texttt{LAMMPS} MD simulations at various temperatures are carried out using the coexisting supercell. The initial temperatures run have only a single MD simulation and there is a large temperature step between the temperatures so that an estimate of where the melting temperature can be ascertained. Additional temperatures can executed to determine the melting temperature. 

In order to establish the melting temperature, there must be at least two temperatures that have found solids and liquids coexistence, with a minimum of four MD simulations run at each temperature. Additionally, there must be a temperature preceding and proceeding the temperatures with only solid or liquid distributions that have at least four MD simulations. These distributions are required in order to get the statistic probability for the solid and liquid phases, from which the results are analyzed and the melting temperature is obtained. The melting temperature fitting procedure is same as it in the original \texttt{SLUSCHI} approach with the \texttt{VASP ab initio MD} simulations \cite{hong_user_2016}.

\section{Results and discussion}

More than two dozen materials (simple and transition metals,  their binary alloys, oxides and carbide) have been  chosen as examples to demonstrate the computational efficiency of our approach. The only factor to consider for the chosen materials is the availability of their \texttt{LASP} potentials.  For a few cases that there are more than one potential available we have also tested  the different potentials for a comparison. The accuracy of the melting temperature results and computational efficiency (i.e., CPU time required to obtain the melting temperature for a material) are the two main aspects we focus on. Our results presented below  are organized into separate sections detailing the outcomes in terms for the melting temperature and the CPU time requirements.

\subsubsection{Computational Time}

\begin{figure*}
    \centering
   \includegraphics[width=0.98\textwidth]{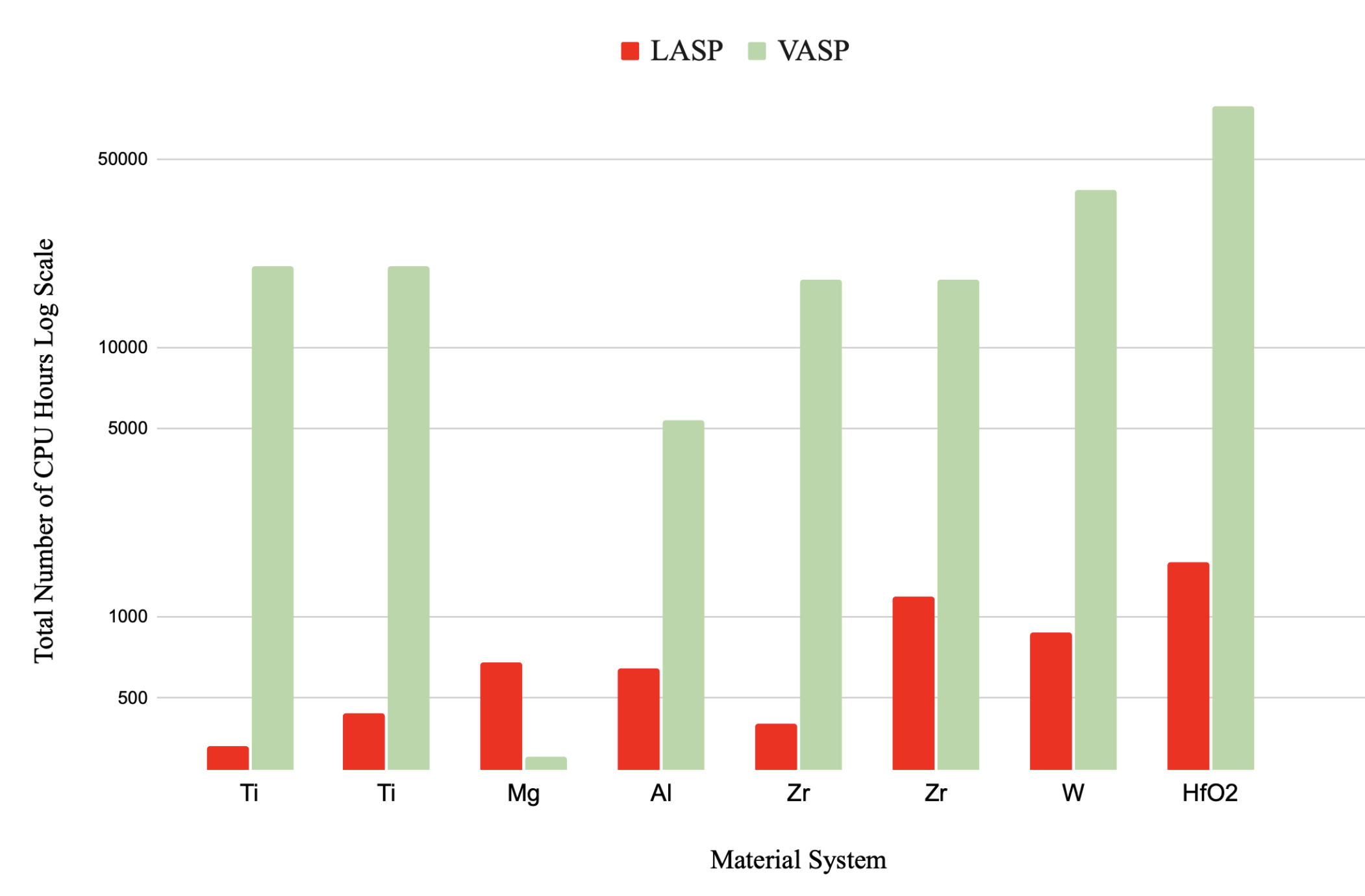}
    \caption{A representation of the total CPU time required for the \texttt{LASP} and original \texttt{SLUSCHI VASP} computational methods. The red being the \texttt{LASP} calculations done with \texttt{LAMMPS}, the green being the data from \texttt{VASP} calculations. }
    \label{fig:LASP_CPU}
\end{figure*}

The chosen materials, including simple metals (Mg, Al), transition metals (Ti, Fe, Ni, Cu, Zr, Mo, Pd, Ag, W, Pt),  binary alloys (AgPd, Ni\_0.\_9Mo\_0.\_1, TiAl3), oxides (TiO\_2, FeO, Fe\_2O\_3, Fe\_3O\_4, NiO, Y\_2O\_3, ZrO\_2, HfO\_2), and carbide (WC), have been studied by the \texttt{SLUSCHI} method interfaced with \texttt{LAMMPS} utilizing the machine learning potentials from \texttt{LASP}. 
In Table \ref{tab:total_lasp_1} we list the CPU times required for the material melting temperature prediction with comparison to available \texttt{DFT} data, along with computational details. For Ti, Ni, Cu, Pd and Pt, their melting temperatures have been calculated with two different \texttt{LASP} potentials.  
The \texttt{LASP} potential database created by Liu et al.\cite{lasp_1,lasp_2} provides a large number of ML potentials with different combinations of elements in each potential. This allows us to explore a broad selection of ML potentials tailored for various element combinations.
Fig. \ref{fig:LASP_CPU} also shows the computational efficiency of \texttt{SLUSCHI} + \texttt{LAMMPS} calculations in comparison to regular \texttt{VASP}-based calculations. From Table \ref{tab:total_lasp_1} and Fig. \ref{fig:LASP_CPU} it is clearly shown that the \texttt{LASP} machine learning potential approach markedly outperforms the \texttt{VASP}-based \texttt{SLUSCHI} calculations, by at least one order of magnitude, in the majority of cases. 
Significant acceleration of material melting temperature predictions, highlighting the effectiveness of the \texttt{SLUSCHI} method with \texttt{LASP} machine learning potentials in enhancing simulation speed, can enable fast screening of numerous candidate materials and pave the way for material rational design within the framework of the materials genome paradigm.

\begin{table}
    \centering
    \caption{Comparison of melting temperatures calculated from \texttt{LASP} machine learning potentials with the \texttt{VASP} and experimental values. The \texttt{VASP} results were reported in Ref. xxx and the experimental values from Ref. xxx. All temperatures are in K. }
    \label{tab:total_lasp_1}
    \begin{tabular}{lrrrrl}
    \hline
     Systems &  $T_m^\mathrm{ML}$ &  $T_m^\mathrm{Corrected}$ & $T_m^\mathrm{VASP}$  & $T_m^\mathrm{exp}$ & ML Potential \\
    \hline
Ag	&$	1326	\pm	19$	&	1302	&	-	&	1208	&	PdAgTiOH	\\
AgPd	&$	1875	\pm	19$	&	1913	&	-	&	1566	&	PdAgTiOH	\\
Al	&$	1170	\pm	11$	&	-	&	1040	&	933	&	AlMgTi	\\
Cu	&$	1050	\pm	31$	&	-	&	-	&	1358	&	PdCuAg	\\
Cu	&$	1130	\pm	5$	&	-	&	-	&	1358	&	CuAlOH	\\
Fe	&$	1774	\pm	17$	&	-	&	-	&	1811	&	FeCr	\\
Fe$_2$O$_3$	&$	1649	\pm	13$	&	1574	&	-	&	1856	&	FeOH	\\
Fe$_3$O$_4$	&$	1718	\pm	5$	&	-	&	-	&	1867	&	FeOH	\\
FeO	&$	1750	\pm	13$	&	-	&	-	&	1633	&	FeOH	\\
HfO$_2$	&$	2986	\pm	29$	&	-	&	-	&	3031	&	SiHfOH	\\
Mg	&$	1186	\pm	13$	&	-	&	-	&	923	&	AlMgTi	\\
Mo	&$	2656	\pm	23$	&	2780	&	-	&	2782	&	PtNiMoOH	\\
Ni	&$	1703	\pm	7$	&	1745	&	-	&	1752	&	PtNiMoOH	\\
Ni	&$	1709	\pm	6$	&	-	&	-	&	1752	&	PdNiCHO	\\
Ni$_{0.9}$Mo$_{0.1}$	&$	1842	\pm	13$	&	1914	&	-	&	1728	&	PtNiMoOH	\\
NiO	&$	2509	\pm	9$	&	-	&	-	&	2228	&	NiO	\\
Pd	&$	1830	\pm	5$	&	1850	&	-	&	1810	&	PdAgTiOH	\\
Pd	&$	1811	\pm	22$	&	1784	&	-	&	1810	&	PdNiCHO	\\
Pt	&$	1805	\pm	5$	&	1930	&	-	&	2090	&	PtNiMoOH	\\
Pt	&$	1863	\pm	9$	&	-	&	-	&	2090	&	PtSnSiO	\\
Ti	&$	1650	\pm	14$	&	1934	&	1750	&	1957	&	PdAgTiOH	\\
Ti	&$	1801	\pm	19$	&	1912	&	1750	&	1957	&	TiOH	\\
TiAl$_3$	&$	1933	\pm	15$	&	-	&	-	&	1673	&	AlMgTi	\\
TiO$_2$	&$	1969	\pm	6$	&	2063	&	-	&	2093	&	TiOH	\\
W	&$	3613	\pm	19$	&	-	&	3497	&	3695	&	WCH	\\
WC	&$	3204	\pm	12$	&	-	&	-	&	3145	&	WCH	\\
Y$_2$O$_3$	&$	2790	\pm	27$	&	-	&	-	&	2718	&	PtYO	\\
Zr	&$	1950	\pm	18$	&	2403	&	2114	&	2128	&	ZrOH	\\
ZrO$_2$	&$	3013	\pm	35$	&	-	&	-	&	2988	&	ZrOH	\\
ZrO$_2$	&$	2685	\pm	61$	&	-	&	-	&	2988	&	ZrO	\\
\hline
RMSE	&	187	&	179	&	151	\\		
\hline
    \bottomrule
    \end{tabular}
\end{table}
\begin{table*}
    \centering
    \caption{Computational details for the LASP simulations, including: Radius of the simulation cell and, number of atoms in the simulation. As well as a the CPU times for melting temperature prediction with the LASP method, and the DFT method.(For the Ti samples the different values for the DFT CPU column come different simulations. (*) one with the best melting temperature and (**) one with the best simulation time.  }
    \label{tab:total_lasp_1}
    \begin{tabular}{lcrccl}
    \hline
     System & Radius (\AA) & N  & LASP & DFT   & LASP   \\
        & & & CPU Hours & CPU Hours & Potential \\    
    \hline
Ag 	&	17 &	512 &	1,205 &	- &	PdAgTiOH	\\				
AgPd	&	16	&	408	&	1,636	&	-	&	PdAgTiOH	\\				
Al	&	16	&	512	&	645	&	5,400	&	AlMgTi	\\				
Cu	&	10	&	192	&	3,643	&	-	&	PdCuAg	\\	
Cu	&	10	&	192	&	5,458	&	-	&	CuAlOH	\\				
Fe	&	10	&	208	&	1,606	&	-	&	FeCr	\\				
Fe$_2$O$_3$	&	16	&	960	&	812	&	-	&	FeOH	\\				
Fe$_3$O$_4$	&	16	&	896	&	1,628	&	-	&	FeOH	\\				
FeO	&	16	&	1,024	&	1,016	&	-	&	FeOH	\\				
HfO$_2$	&	15	&	648	&	1,588	&	7,900	&	SiHfOH	\\				
Mg	&	16	&	360	&	679	&	304	&	AlMgTi	\\				
Mo	&	16	&	500	&	423	&	-	&	PtNiMoOH	\\				
Ni	&	16	&	800	&	992	&	-	&	PtNiMoOH	\\				
Ni	&	16	&	800	&	582	&	-	&	PdNiCHO	\\				
Ni$_{0.9}$Mo$_{0.1}$	&	14	&	576	&	532	&	-	&	PtNiMoOH	\\			
NiO	&	16	&	928	&	1,034	&	-	&	NiO	\\				
Pd	&	16	&	512	&	919	&	-	&	PdAgTiOH	\\				
Pd	&	16	&	512	&	459	&	-	&	PdNiCHO	\\				
Pt	&	16	&	512	&	230	&	-	&	PtNiMoOH	\\				
Pt	&	16	&	512	&	435	&	-	&	PtSnSiO	\\				
Ti	&	16	&	500	&	697	&	20,000*	&	PdAgTiOH	\\	
Ti	&	16	&	500	&	694	&	7,900**	&	TiOH	\\	
TiAl$_3$	&	15	&	512	&	1,391	&	-	&	AlMgTi	\\				
TiO$_2$	&	17	&	780	&	3,122	&	-	&	TiOH	\\				
W	&	10	&	132	&	869	&	35,900	&	WCH	\\				
WC	&	15	&	696	&	909	&	-	&	WCH	\\				
Y$_2$O$_3$	&	16	&	600	&	1,380	&	-	&	PtYO	\\				
Zr	&	15	&	336	&	402	&	17,841	&	ZrOH	\\				
ZrO$_2$	&	10	&	192	&	1,703	&	-	&	ZrOH	\\				
ZrO$_2$	&	10	&	192	&	1,558	&	-	&	ZrO	\\				

    \hline
    \bottomrule
    \end{tabular} 
\end{table*}

\subsubsection{Melting Temperature}

\begin{figure}
    \centering
    \includegraphics[width=0.49\textwidth]{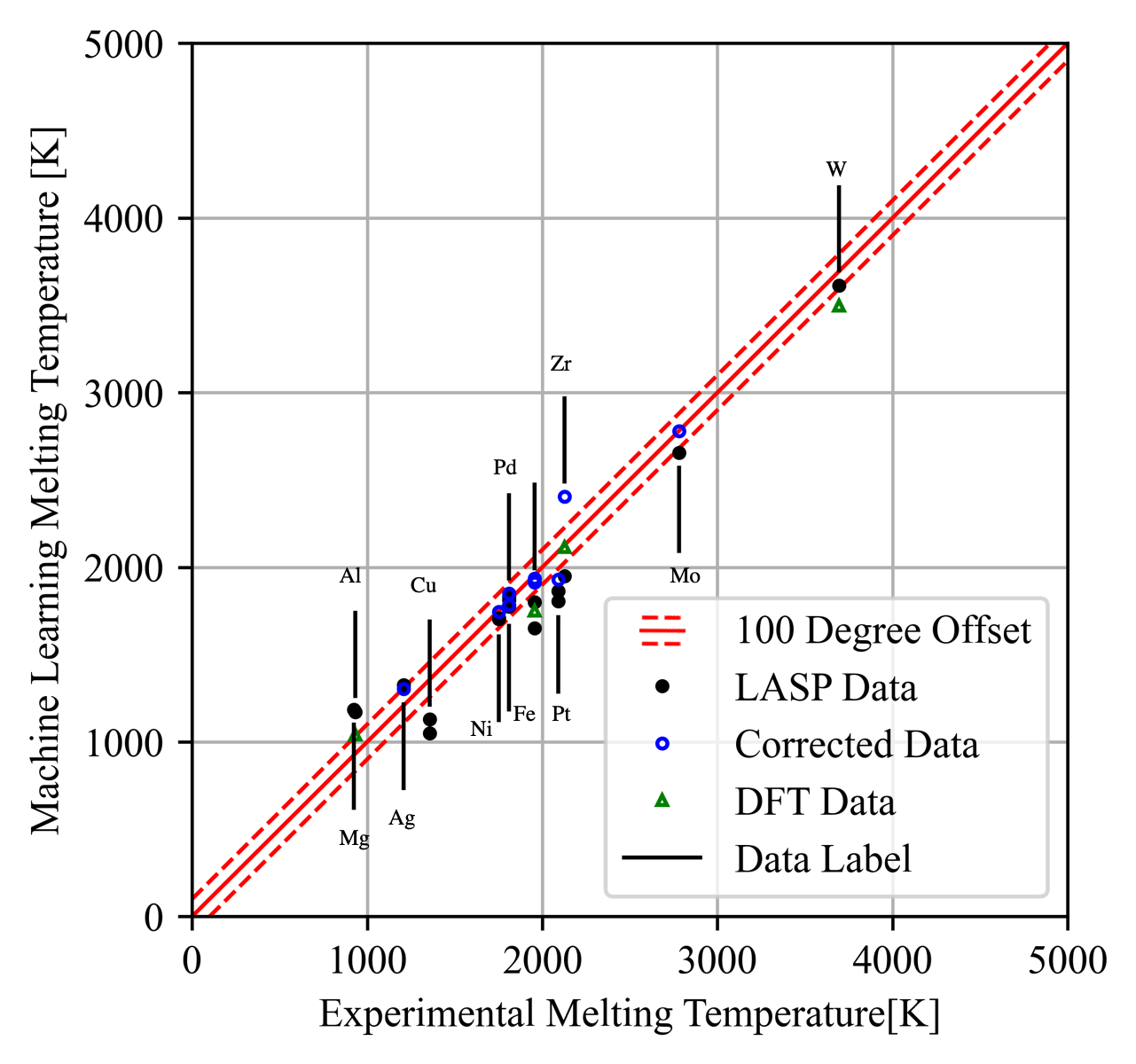}
    \caption{The LASP, corrected and DFT data graphed against the experimental melting temperatures for single element calculations. The solid diagonal is perfect agreement between the data sets and the two dashed lines are 100K off sets of perfect agreement. The black vertical lines indicate the elements present for the 3 data sets shown. The full circles are the raw LASP data, the open circles represent the corrected LASP data, and the open triangles represent the DFT results. The root mean square (RMSE) value for the raw data is 188, while the corrected data has an RMSE value of 113. Both of which are close to the RMSE value for previously proven DFT calculation techniques.
}
    \label{fig:lammps_corrected}
\end{figure}

\begin{figure}
    \centering
    \includegraphics[width=0.49\textwidth]{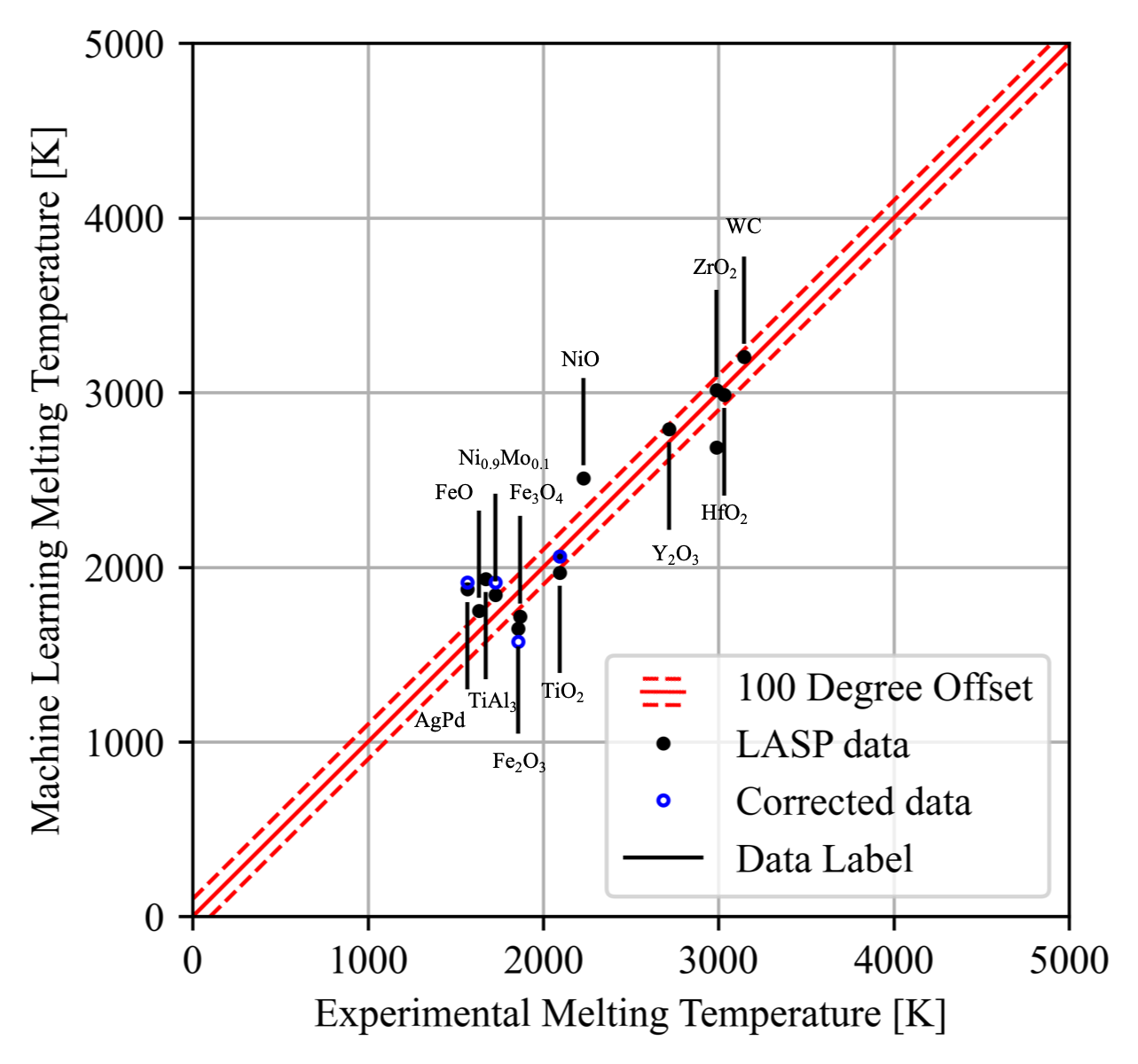}
    \caption{The comparison between the experimental melting temperature and the original LASP data and the corrected LASP data. The solid diagonal is perfect agreement between the data sets and the two dashed lines are 100K off sets of perfect agreement. The black vertical lines indicate the elements present for the 3 data sets shown. The full circles are the original LASP data, the open circles represent the corrected LASP data. The RMSE value for the raw data is 159, while the corrected data has an RMSE value of 246.  }
    \label{fig:LAMMPS_compound_corrected}
\end{figure}

Although a dramatic reduction in the CPU time is achieved the accuracy of the calculated melting temperatures is another key aspect needed to be addressed.
Evaluating the precision of melting temperatures is paramount to decide the applicability of  our implementation of  \texttt{LAMMPS} with the \texttt{LASP} machine learning potentials in the \texttt{SLUSCHI} code.  Table \ref{tab:total_lasp_1} summarizes the melting temperatures derived using \texttt{LASP} with \texttt{LAMMPS} together with the corrected \texttt{LASP} data. The \texttt{VASP} DFT data and experimental values are also included in Table \ref{tab:total_lasp_1} for comparison. 
Fig. \ref{fig:lammps_corrected} and  Fig. \ref{fig:LAMMPS_compound_corrected} 
show the results for single element and compound systems, respectively. 
Analysis shows that 33\% of the melting temperatures are within 100 K of experimental values for the combined single element and compound data sets. The number increase to 60\% when the margin of error is increased to within 200 K of experimental values. The data sets have a root mean square value (RMSE) of 187 K for the raw data in contrast to 151 K of \texttt{VASP} DFT data. The root mean square value difference is 36 K, indicating that  the melting temperatures predicted using \texttt{LASP} with \texttt{LAMMPS} is reasonably accurate compared to the DFT results. 

In order to determine the melting temperature of a material in close to the accuracy of the \texttt{VASP}-based \texttt{SLUSCHI} approach, we have conducted a correction based on the heat of fusion.  
The correction within the dataset employs a straightforward approach, linking the melting temperature ($T_m$) to the heat of fusion ($\Delta H$) for each element, as depicted in Eqn. \ref{DFT_Correction}. This relationship allows for recalculating the projected melting temperatures from DFT calculations using the ML-derived heat of fusion and the heat of fusion from conventional DFT calculations. Given that \texttt{LAMMPS} reports the energy for each simulation step, determining the $\Delta H$ for $T_m^\mathrm{ML}$ is as straightforward as applying an ensemble average from selected material system snapshots, illustrated in Eqn. \ref{DFT_correction_2}.

\begin{equation}
\frac{T_m^\mathrm{DFT}}{T_m^\mathrm{ML}} = \frac{\Delta H^\mathrm{DFT}}{\Delta H^\mathrm{ML}}
\label{DFT_Correction}
\end{equation}

\begin{equation}
H^\mathrm{DFT} - H^\mathrm{ML} = \left\langle H^\mathrm{DFT} - H^\mathrm{ML}\right\rangle_{H^\mathrm{ML}}
\label{DFT_correction_2}
\end{equation}

An example calculation is provided in Table \ref{tab:example_calculation}, detailing how this correction is applied. Table \ref{tab:example_calculation} outlines the system's specifics, including temperature and run number, alongside the corresponding energy data. From these, the heat of fusion is calculated for both liquid and solid states, and the ratio from Eqn. \ref{DFT_Correction} determines the adjusted melting temperature for DFT-based predictions. In the case of Ti, this correction led to an increased temperature estimate, bring the final temperature closer to experimental values. 

Post-correction indicates an improvement in accuracy for the single element calculations, resulting in the majority of corrected values more closely aligning with the DFT values and experimental observations. The overall RMSE value for both the single element and compound calculations after corrections is 179 K, which is comparable to the value of 187 K for the uncorrected raw data.  We notice that the nearly unchanged RMSE value for the corrected data is mainly due to the worse melting temperature results for alloys and oxides after correction compared to the uncorrected ones.
 Therefore, we recommend the correction for the single element calculations and it improves the accuracy of the melting temperature results. Overall, the \texttt{SLUSCHI} + \texttt{LASP} approach can be applied for a quick screening of candidate materials in a reasonable accuracy of the melting temperature predictions.

\begin{table}
    \centering
    \caption{Melting Temperature DFT Data Correction Calculation}
    \label{tab:example_calculation}
    \begin{tabular}{lcrr}
    \hline
    & \multicolumn{1}{c}{Potential PdAgTiOH} \\
    \cmidrule{1-3}
    \hline
    Material System & Ti  \\
    \hline
    Liquid & & & \\
    \hline
    Run & LASP & VASP \\
    \hline
    1600\_4 & -3716.89 & -3684.23\\
    1650\_1 & -3712.92 & -3680.20\\
    1675\_3 & -3705.83 & -3668.19\\
    1675\_4 & -3704.26 & -3669.26\\
    \hline
    Average & -3709.98 & -3675.47 \\
    \hline
    Solid & & \\
    \hline
     Run & LASP & VASP \\
    \hline
    1600\_1 & -3765.32 & -3740.32 \\
    1600\_2 & -3775.89 & -3752.78 \\
    1650\_3 & -3762.90 & -3737.35 \\
    1650\_4 & -3772.13 & -3748.40 \\
    
    \hline
    Average & -3769.06 & -3744.71 \\
    \hline
    Heat of Fusion & 59.09 & 69.24 \\
    \hline
    LASP + LAMMPS MT & \multicolumn{2}{c}{1650} \\
    \hline
    DFT MT & \multicolumn{2}{c}{1934} \\
    \bottomrule
    \end{tabular}
\end{table}


\section{Conclusions}
The \texttt{SLUSCHI} package is implemented to interface with the highly  efficient molecular dynamics \texttt{LAMMPS} which can run the simulations for material melting temperature predictions with the machine learning \texttt{LASP} potentials. 
The \texttt{LASP} potentials constructed by  integrating the neural network potential technique with the global potential energy surface exploration method
 from the data (energies, forces and stresses) of first-principles calculations, and have been widely applied for structure prediction and reaction mechanism exploration. 
This newly implemented approach has been tested to predict the  melting temperatures for various materials, including simple and transition metals, alloys, oxides and carbide. 
Because the computationally expensive \texttt{AIMD VASP} calculations are completely eliminated in the melting temperature prediction the dramatic speed-up and CPU time reduction 
is achieved by the present approach. The CPU time results indicate that the \texttt{SLUSCHI} + \texttt{LAMMPS} approach employing the \texttt{LASP} potentials outperforms the original  \texttt{VASP}-based approach by at least one order of magnitude in the majority of cases.
The calculated melting temperatures for the chosen materials (simple and transition metals, alloys, oxides and carbide) are reasonably accurate with the RMSE value of 187 K for the uncorrected results, compared to 151 K for the data obtained completely from first-principles calculations. We also show that  the post-correction can improve the accuracy of 
melting temperatures for the single elemental materials.  
The present study demonstrates that the new implementation interfacing the \texttt{LAMMPS} molecular dynamics with the \texttt{SLUSCHI} package is very useful for screening the candidate materials and facilitates the realization of rational material design within the materials genome paradigm. 
 
\bibliography{iron_melting_curve/bibliography}

\begin{thebibliography}{18}%
\makeatletter
\providecommand \@ifxundefined [1]{%
 \@ifx{#1\undefined}
}%
\providecommand \@ifnum [1]{%
 \ifnum #1\expandafter \@firstoftwo
 \else \expandafter \@secondoftwo
 \fi
}%
\providecommand \@ifx [1]{%
 \ifx #1\expandafter \@firstoftwo
 \else \expandafter \@secondoftwo
 \fi
}%
\providecommand \natexlab [1]{#1}%
\providecommand \enquote  [1]{``#1''}%
\providecommand \bibnamefont  [1]{#1}%
\providecommand \bibfnamefont [1]{#1}%
\providecommand \citenamefont [1]{#1}%
\providecommand \href@noop [0]{\@secondoftwo}%
\providecommand \href [0]{\begingroup \@sanitize@url \@href}%
\providecommand \@href[1]{\@@startlink{#1}\@@href}%
\providecommand \@@href[1]{\endgroup#1\@@endlink}%
\providecommand \@sanitize@url [0]{\catcode `\\12\catcode `\$12\catcode
  `\&12\catcode `\#12\catcode `\^12\catcode `\_12\catcode `\%12\relax}%
\providecommand \@@startlink[1]{}%
\providecommand \@@endlink[0]{}%
\providecommand \url  [0]{\begingroup\@sanitize@url \@url }%
\providecommand \@url [1]{\endgroup\@href {#1}{\urlprefix }}%
\providecommand \urlprefix  [0]{URL }%
\providecommand \Eprint [0]{\href }%
\providecommand \doibase [0]{http://dx.doi.org/}%
\providecommand \selectlanguage [0]{\@gobble}%
\providecommand \bibinfo  [0]{\@secondoftwo}%
\providecommand \bibfield  [0]{\@secondoftwo}%
\providecommand \translation [1]{[#1]}%
\providecommand \BibitemOpen [0]{}%
\providecommand \bibitemStop [0]{}%
\providecommand \bibitemNoStop [0]{.\EOS\space}%
\providecommand \EOS [0]{\spacefactor3000\relax}%
\providecommand \BibitemShut  [1]{\csname bibitem#1\endcsname}%
\let\auto@bib@innerbib\@empty
\bibitem [{\citenamefont {Hong}\ and\ \citenamefont {van~de
  Walle}()}]{hong_user_2016}%
  \BibitemOpen
  \bibfield  {author} {\bibinfo {author} {\bibfnamefont {Q.-J.}\ \bibnamefont
  {Hong}}\ and\ \bibinfo {author} {\bibfnamefont {A.}~\bibnamefont {van~de
  Walle}},\ }\href {\doibase 10.1016/j.calphad.2015.12.003} {\bibfield
  {journal} {\bibinfo  {journal} {Calphad}\ }\textbf {\bibinfo {volume} {52}},\
  \bibinfo {pages} {88}}\BibitemShut {NoStop}%
\bibitem [{\citenamefont {Kirkwood}()}]{Free_energy_3}%
  \BibitemOpen
  \bibfield  {author} {\bibinfo {author} {\bibfnamefont {J.~G.}\ \bibnamefont
  {Kirkwood}},\ }\href {\doibase 10.1063/1.1749657} {\bibfield  {journal}
  {\bibinfo  {journal} {The Journal of Chemical Physics}\ }\textbf {\bibinfo
  {volume} {3}},\ \bibinfo {pages} {300}}\BibitemShut {NoStop}%
\bibitem [{\citenamefont {Lin}\ \emph {et~al.}()\citenamefont {Lin},
  \citenamefont {Blanco},\ and\ \citenamefont {Goddard}}]{Free_energy_4}%
  \BibitemOpen
  \bibfield  {author} {\bibinfo {author} {\bibfnamefont {S.-T.}\ \bibnamefont
  {Lin}}, \bibinfo {author} {\bibfnamefont {M.}~\bibnamefont {Blanco}}, \ and\
  \bibinfo {author} {\bibfnamefont {W.~A.}\ \bibnamefont {Goddard}},\ }\href
  {\doibase 10.1063/1.1624057} {\bibfield  {journal} {\bibinfo  {journal} {The
  Journal of Chemical Physics}\ }\textbf {\bibinfo {volume} {119}},\ \bibinfo
  {pages} {11792}}\BibitemShut {NoStop}%
\bibitem [{\citenamefont {Widom}()}]{Free_energy_5}%
  \BibitemOpen
  \bibfield  {author} {\bibinfo {author} {\bibfnamefont {B.}~\bibnamefont
  {Widom}},\ }\href {\doibase 10.1063/1.1734110} {\bibfield  {journal}
  {\bibinfo  {journal} {The Journal of Chemical Physics}\ }\textbf {\bibinfo
  {volume} {39}},\ \bibinfo {pages} {2808}}\BibitemShut {NoStop}%
\bibitem [{\citenamefont {van~de Walle}\ and\ \citenamefont
  {Ceder}()}]{Free_energy_method_1}%
  \BibitemOpen
  \bibfield  {author} {\bibinfo {author} {\bibfnamefont {A.}~\bibnamefont
  {van~de Walle}}\ and\ \bibinfo {author} {\bibfnamefont {G.}~\bibnamefont
  {Ceder}},\ }\href {\doibase 10.1103/RevModPhys.74.11} {\bibfield  {journal}
  {\bibinfo  {journal} {Reviews of Modern Physics}\ }\textbf {\bibinfo {volume}
  {74}},\ \bibinfo {pages} {11}}\BibitemShut {NoStop}%
\bibitem [{\citenamefont {Kofke}\ and\ \citenamefont
  {Cummings}()}]{Free_energy_method_2}%
  \BibitemOpen
  \bibfield  {author} {\bibinfo {author} {\bibfnamefont {D.~A.}\ \bibnamefont
  {Kofke}}\ and\ \bibinfo {author} {\bibfnamefont {P.~T.}\ \bibnamefont
  {Cummings}},\ }\href {\doibase 10.1080/002689797169600} {\bibfield  {journal}
  {\bibinfo  {journal} {Molecular Physics}\ }\textbf {\bibinfo {volume} {92}},\
  \bibinfo {pages} {973}}\BibitemShut {NoStop}%
\bibitem [{\citenamefont {Alfè}(2009)}]{large_scale_1}%
  \BibitemOpen
  \bibfield  {author} {\bibinfo {author} {\bibfnamefont {D.}~\bibnamefont
  {Alfè}},\ }\href {\doibase 10.1103/PhysRevB.79.060101} {\bibfield  {journal}
  {\bibinfo  {journal} {Physical Review B}\ }\textbf {\bibinfo {volume} {79}},\
  \bibinfo {pages} {060101} (\bibinfo {year} {2009})}\BibitemShut {NoStop}%
\bibitem [{\citenamefont {Mei}\ and\ \citenamefont
  {Davenport}()}]{large_scale_2}%
  \BibitemOpen
  \bibfield  {author} {\bibinfo {author} {\bibfnamefont {J.}~\bibnamefont
  {Mei}}\ and\ \bibinfo {author} {\bibfnamefont {J.~W.}\ \bibnamefont
  {Davenport}},\ }\href {\doibase 10.1103/PhysRevB.46.21} {\bibfield  {journal}
  {\bibinfo  {journal} {Physical Review B}\ }\textbf {\bibinfo {volume} {46}},\
  \bibinfo {pages} {21}}\BibitemShut {NoStop}%
\bibitem [{\citenamefont {Belonoshko}\ \emph
  {et~al.}({\natexlab{a}})\citenamefont {Belonoshko}, \citenamefont
  {Skorodumova}, \citenamefont {Rosengren},\ and\ \citenamefont
  {Johansson}}]{Fast_heating_1}%
  \BibitemOpen
  \bibfield  {author} {\bibinfo {author} {\bibfnamefont {A.~B.}\ \bibnamefont
  {Belonoshko}}, \bibinfo {author} {\bibfnamefont {N.~V.}\ \bibnamefont
  {Skorodumova}}, \bibinfo {author} {\bibfnamefont {A.}~\bibnamefont
  {Rosengren}}, \ and\ \bibinfo {author} {\bibfnamefont {B.}~\bibnamefont
  {Johansson}},\ }\href {\doibase 10.1103/PhysRevB.73.012201} {\bibfield
  {journal} {\bibinfo  {journal} {Physical Review B}\ }\textbf {\bibinfo
  {volume} {73}},\ \bibinfo {pages} {012201} ({\natexlab{a}})}\BibitemShut
  {NoStop}%
\bibitem [{\citenamefont {Alfè}\ \emph {et~al.}()\citenamefont {Alfè},
  \citenamefont {Cazorla},\ and\ \citenamefont {Gillan}}]{Fast_heating_2}%
  \BibitemOpen
  \bibfield  {author} {\bibinfo {author} {\bibfnamefont {D.}~\bibnamefont
  {Alfè}}, \bibinfo {author} {\bibfnamefont {C.}~\bibnamefont {Cazorla}}, \
  and\ \bibinfo {author} {\bibfnamefont {M.~J.}\ \bibnamefont {Gillan}},\
  }\href {\doibase 10.1063/1.3605601} {\bibfield  {journal} {\bibinfo
  {journal} {The Journal of Chemical Physics}\ }\textbf {\bibinfo {volume}
  {135}},\ 10.1063/1.3605601},\ \bibinfo {note} {publisher: {AIP}
  Publishing}\BibitemShut {NoStop}%
\bibitem [{\citenamefont {Haskins}\ \emph {et~al.}()\citenamefont {Haskins},
  \citenamefont {Moriarty},\ and\ \citenamefont {Hood}}]{Fast_heating_3}%
  \BibitemOpen
  \bibfield  {author} {\bibinfo {author} {\bibfnamefont {J.~B.}\ \bibnamefont
  {Haskins}}, \bibinfo {author} {\bibfnamefont {J.~A.}\ \bibnamefont
  {Moriarty}}, \ and\ \bibinfo {author} {\bibfnamefont {R.~Q.}\ \bibnamefont
  {Hood}},\ }\href {\doibase 10.1103/PhysRevB.86.224104} {\bibfield  {journal}
  {\bibinfo  {journal} {Physical Review B}\ }\textbf {\bibinfo {volume} {86}},\
  \bibinfo {pages} {224104}}\BibitemShut {NoStop}%
\bibitem [{\citenamefont {Belonoshko}\ \emph
  {et~al.}({\natexlab{b}})\citenamefont {Belonoshko}, \citenamefont {Lukinov},
  \citenamefont {Burakovsky}, \citenamefont {Preston},\ and\ \citenamefont
  {Rosengren}}]{Fast_heating_4}%
  \BibitemOpen
  \bibfield  {author} {\bibinfo {author} {\bibfnamefont {A.~B.}\ \bibnamefont
  {Belonoshko}}, \bibinfo {author} {\bibfnamefont {T.}~\bibnamefont {Lukinov}},
  \bibinfo {author} {\bibfnamefont {L.}~\bibnamefont {Burakovsky}}, \bibinfo
  {author} {\bibfnamefont {D.~L.}\ \bibnamefont {Preston}}, \ and\ \bibinfo
  {author} {\bibfnamefont {A.}~\bibnamefont {Rosengren}},\ }\href {\doibase
  10.1140/epjst/e2013-01743-1} {\bibfield  {journal} {\bibinfo  {journal} {The
  European Physical Journal Special Topics}\ }\textbf {\bibinfo {volume}
  {216}},\ \bibinfo {pages} {199} ({\natexlab{b}})}\BibitemShut {NoStop}%
\bibitem [{\citenamefont {Hong}\ and\ \citenamefont {Van
  De~Walle}(2013)}]{SLUSCHI_table_1}%
  \BibitemOpen
  \bibfield  {author} {\bibinfo {author} {\bibfnamefont {Q.-J.}\ \bibnamefont
  {Hong}}\ and\ \bibinfo {author} {\bibfnamefont {A.}~\bibnamefont {Van
  De~Walle}},\ }\href@noop {} {\bibfield  {journal} {\bibinfo  {journal} {The
  Journal of chemical physics}\ }\textbf {\bibinfo {volume} {139}} (\bibinfo
  {year} {2013})}\BibitemShut {NoStop}%
\bibitem [{\citenamefont {Hong}\ \emph {et~al.}(2015)\citenamefont {Hong},
  \citenamefont {Ushakov}, \citenamefont {Navrotsky},\ and\ \citenamefont {Van
  De~Walle}}]{SLUSCHI_table_2}%
  \BibitemOpen
  \bibfield  {author} {\bibinfo {author} {\bibfnamefont {Q.-J.}\ \bibnamefont
  {Hong}}, \bibinfo {author} {\bibfnamefont {S.~V.}\ \bibnamefont {Ushakov}},
  \bibinfo {author} {\bibfnamefont {A.}~\bibnamefont {Navrotsky}}, \ and\
  \bibinfo {author} {\bibfnamefont {A.}~\bibnamefont {Van De~Walle}},\
  }\href@noop {} {\bibfield  {journal} {\bibinfo  {journal} {Acta Materialia}\
  }\textbf {\bibinfo {volume} {84}},\ \bibinfo {pages} {275} (\bibinfo {year}
  {2015})}\BibitemShut {NoStop}%
\bibitem [{\citenamefont {Hong}\ and\ \citenamefont {van~de
  Walle}(2015)}]{SLUSCHI_table_3}%
  \BibitemOpen
  \bibfield  {author} {\bibinfo {author} {\bibfnamefont {Q.-J.}\ \bibnamefont
  {Hong}}\ and\ \bibinfo {author} {\bibfnamefont {A.}~\bibnamefont {van~de
  Walle}},\ }\href {\doibase 10.1103/PhysRevB.92.020104} {\bibfield  {journal}
  {\bibinfo  {journal} {Physical Review B}\ }\textbf {\bibinfo {volume} {92}},\
  \bibinfo {pages} {020104} (\bibinfo {year} {2015})}\BibitemShut {NoStop}%
\bibitem [{\citenamefont {Miljacic}\ \emph {et~al.}(2015)\citenamefont
  {Miljacic}, \citenamefont {Demers}, \citenamefont {Hong},\ and\ \citenamefont
  {Van De~Walle}}]{SLUSCHI_table_4}%
  \BibitemOpen
  \bibfield  {author} {\bibinfo {author} {\bibfnamefont {L.}~\bibnamefont
  {Miljacic}}, \bibinfo {author} {\bibfnamefont {S.}~\bibnamefont {Demers}},
  \bibinfo {author} {\bibfnamefont {Q.-J.}\ \bibnamefont {Hong}}, \ and\
  \bibinfo {author} {\bibfnamefont {A.}~\bibnamefont {Van De~Walle}},\
  }\href@noop {} {\bibfield  {journal} {\bibinfo  {journal} {Calphad}\ }\textbf
  {\bibinfo {volume} {51}},\ \bibinfo {pages} {133} (\bibinfo {year}
  {2015})}\BibitemShut {NoStop}%
\bibitem [{\citenamefont {Huang}\ \emph {et~al.}(2019)\citenamefont {Huang},
  \citenamefont {Shang}, \citenamefont {Kang}, \citenamefont {Zhang},\ and\
  \citenamefont {Liu}}]{lasp_1}%
  \BibitemOpen
  \bibfield  {author} {\bibinfo {author} {\bibfnamefont {S.-D.}\ \bibnamefont
  {Huang}}, \bibinfo {author} {\bibfnamefont {C.}~\bibnamefont {Shang}},
  \bibinfo {author} {\bibfnamefont {P.-L.}\ \bibnamefont {Kang}}, \bibinfo
  {author} {\bibfnamefont {X.-J.}\ \bibnamefont {Zhang}}, \ and\ \bibinfo
  {author} {\bibfnamefont {Z.-P.}\ \bibnamefont {Liu}},\ }\href@noop {}
  {\bibfield  {journal} {\bibinfo  {journal} {Wiley Interdisciplinary Reviews:
  Computational Molecular Science}\ }\textbf {\bibinfo {volume} {9}},\ \bibinfo
  {pages} {e1415} (\bibinfo {year} {2019})}\BibitemShut {NoStop}%
\bibitem [{\citenamefont {Kang}\ \emph {et~al.}(2021)\citenamefont {Kang},
  \citenamefont {Shang},\ and\ \citenamefont {Liu}}]{lasp_2}%
  \BibitemOpen
  \bibfield  {author} {\bibinfo {author} {\bibfnamefont {P.-l.}\ \bibnamefont
  {Kang}}, \bibinfo {author} {\bibfnamefont {C.}~\bibnamefont {Shang}}, \ and\
  \bibinfo {author} {\bibfnamefont {Z.-p.}\ \bibnamefont {Liu}},\ }\href@noop
  {} {\bibfield  {journal} {\bibinfo  {journal} {Chinese Journal of Chemical
  Physics}\ }\textbf {\bibinfo {volume} {34}},\ \bibinfo {pages} {583}
  (\bibinfo {year} {2021})}\BibitemShut {NoStop}%
\end{thebibliography}%

\end{document}